\documentclass[prl, twocolumn]{revtex4-1}
\usepackage{graphicx}
\begin{document}
\title{Evolution-free Hamiltonian parameter estimation through Zeeman markers}
\author{Daniel Burgarth}
\affiliation{Institute of Mathematics, Physics and Computer Science, Aberystwyth University, Aberystwyth SY23 3BZ, UK}
\author{Ashok Ajoy}
\affiliation{Department of Chemistry, University of California Berkeley and Materials Science Division Lawrence Berkeley National Laboratory, Berkeley CA}
\date{\today}
\begin{abstract}
We provide a protocol for Hamiltonian parameter estimation which relies only on the Zeeman effect. No time-dependent quantities need to be measured, it fully suffices to observe spectral shifts induced by fields applied to local `markers'. We demonstrate the idea with a simple tight-binding Hamiltonian and numerically show stability with respect to Gaussian noise on the spectral measurements. Then we generalize the result to show applicability to a wide range of systems, including quantum spin chains, networks of qubits, and coupled harmonic oscillators, and suggest potential experimental implementations.
\end{abstract}
\maketitle
\textit{Introduction.---}
One of the most fundamental concepts of Quantum Theory is the Hamiltonian as the generator of dynamics. Hamiltonians play a paramount importance in our understanding of matter and its properties, but they can also do some work for us: in quantum technology, they can trigger quantum simulations or even form the basis of quantum computing. The current drive towards high-fidelity quantum devices has kindled renewed interest in Hamiltonian parameter estimation \cite{jared}.

Hamiltonian parameter estimation is important but costly in terms of resources used. It requires detailed control and measurements, as well as extensive post-processing of the measured data. In specific situations, however, specialized  protocols can ease the task. As such, \emph{indirect} estimation has recently been developed \cite{couplingestimation} to allow parameter estimation in systems with limited access.
The basic idea is to measure a free induction decay of the type $f(t)=\langle 1|\exp{(-iHt)}|1\rangle=\sum_k \exp{(-ie_kt)|\langle 1|e_k\rangle|^2}$ with respect to some reference state $|1\rangle$ representing a `local' probe. Fourier transform then provides the spectrum $\{e_k\}$ and the coefficients $|\langle 1|e_k\rangle|$, and it was shown that for one-dimensional systems this limited data can suffice to estimate the \emph{full} system.

This initial work has been extended in several directions. It was shown \cite{carlo}, perhaps surprisingly, that for certain systems the initialization in a reference state is not necessary. If more than one probe can be used, the method can be applied to arbitrary networks \cite{infecting}. Other physical systems, such as fermionic and bosonic networks \cite{quadratic} and linear passive systems \cite{guta} were also found to be indirectly estimable. It was found that additional control of the probes helps to gain phase information \cite{qsi} and a description in terms of polynomial equations were developed in \cite{sarovar, sarovarext, paola}. Graph structures occuring in biometric systems were studied in \cite{birgitta} and the first NMR experiment demonstrated feasibility of the method for small systems \cite{elham, experiment}.

All these methods suffer from a major drawback: the requirement to measure time-resolved dynamics and to do so locally. This is hard, because the dynamics can be very fast for strongly coupled system; because decoherence limits the time period over which useful data can be acquired, and because local measurements are difficult.
It is interesting to note, however, that half of the required data for the above schemes, namely the spectrum $\{e_k\}$, can be relatively easy to measure. The spectrum is a global property of a many-body system and can be measured by absorption or emission of electromagnetic radiation, and many advanced methods for spectroscopy in a plethora of experiments have been established. Generally spectroscopy does not require time-resolved measurements and can work well in the presence of decoherence.
\begin{figure}
\includegraphics[width=0.45\textwidth]{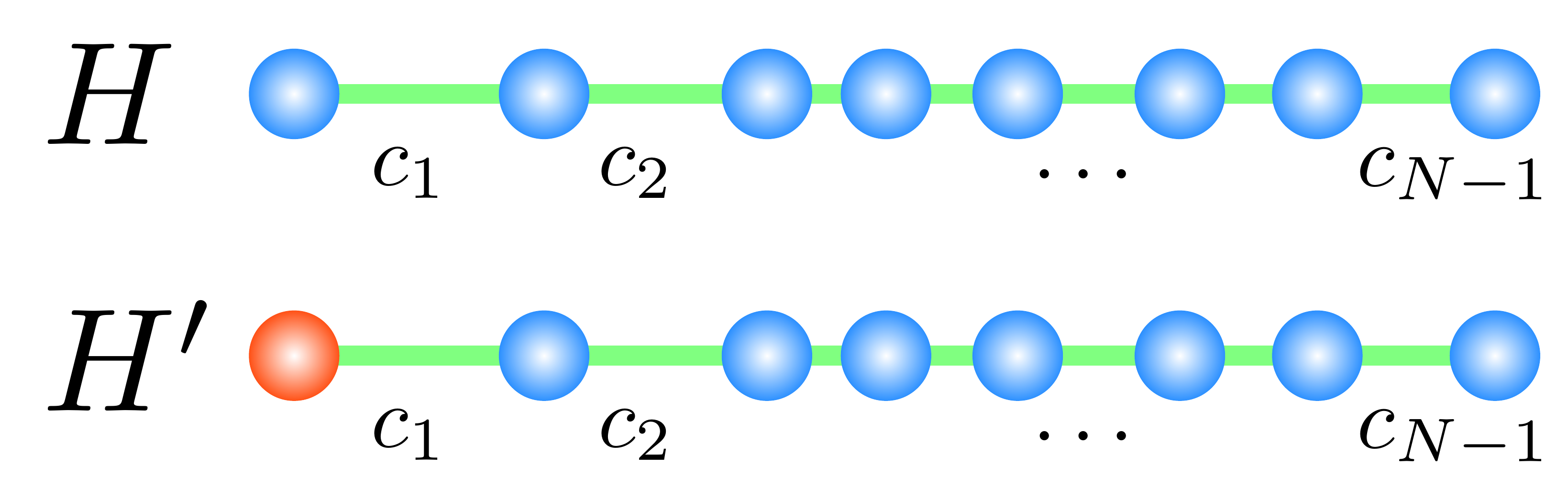}
\caption{We estimate the parameters $c_n$ of a Hamiltonian $H$ given by Eq.~(\ref{hamiltonian}) by actively modifying it to $H'$, applying a time-independent local field on the `Zeeman marker' (orange). Performing spectroscopy on $H$ and $H'$ yields two spectra $\{e_k\}$ and $\{e'_n\}$, from which we compute the $c_n$ through Eqs.~(\ref{mainequation}),(\ref{polynomial}).\label{chain}}
\end{figure}
Does the spectrum provide enough information to perform Hamiltonian parameter estimation? As can be seen from \cite{couplingestimation}, one fixed spectrum can give rise to infinitely many different parameter choices (through fixing $\langle 1 |e_k \rangle$). However, we show in here that \emph{two spectra} do the job: one being the original spectrum of $H$, and one being the spectrum of a modified Hamiltonian $H'$ which arrises from applying a local field to a probe which we call `Zeeman marker'. This probe does not have to be measured locally and no time-dependent data is required. We first demonstrate this idea with a simple (but common) tight binding Hamiltonian, then analyze its stability, and finally generalize to spin Hamiltonians, free fermions and bosons, and arbitrary networks.

\textit{Simple Model.---}
We consider a $N-$dimensional Hilbert space with basis $\{|1\rangle,|2\rangle,\ldots,|N\rangle\}$ and a tight-binding Hamiltonian given by
\begin{equation}
\label{hamiltonian}
	H=\sum_{n=1}^{N-1} c_n |n\rangle\langle n+1|+c_n^* |n+1\rangle\langle n|.
\end{equation}
Although in general the parameters $c_n$ could be complex, it suffices to consider the case $c_n>0$, because complex phases $e^{i \phi_n}$ can always be absorbed in the choice of basis,
$\{|2\rangle\rightarrow e^{-i\phi_1}|2\rangle,|3\rangle\rightarrow e^{-i(\phi_2+\phi_1)}|3\rangle,\cdots,|N\rangle\rightarrow e^{-i(\phi_{N-1}+\phi_{N-2})}|N\rangle\}$. Our target is to estimate the parameters $c_n$ by measuring the spectrum of the system only.
In order to do so, let us apply a field $f|1\rangle\langle1|$ at site 1 (see Fig.~\ref{chain}. Thus, we have modified our Hamiltonian to $H'=H+f|1\rangle\langle1|$. Denoting the spectrum of the original Hamiltonian $H$ as $\{e_k\}$ and the one of $H'$ as $\{e'_n\}$ we can derive (see the appendix)
\begin{equation}
		|\langle e_k|1\rangle |^2=(e_k'-e_k)/f\prod_{m\neq k}\frac{(e_k-e'_m)}{(e_k-e_m)}
		\label{mainequation}
\end{equation}
Without loss of generality we may choose the phases of the eigenstates $|e_k\rangle$ such that $\langle e_k | 1 \rangle \ge 0$. This means that the measurements of the spectra of $H$ and $H'$ reveal $\langle e_k | 1 \rangle$. We remark that the value of $f$ plays no significant role and indeed can be unknown: the $|\langle e_k|1\rangle |^2$ sum up to $1$ which implies that $f$ can also be inferred from the spectra.

From $\langle e_k | 1\rangle$ we can obtain the $c_n$ following \cite{couplingestimation, sarovar}: the equation
\begin{equation}
\label{polynomial}
	\sum_{k=1}^N e_k^m |\langle 1 | e_k \rangle|^2= \langle 1 |H^m |1\rangle
\end{equation}
provides iterative polynomial equations in the $c_n$ which can be solved, e.g. for $m=2$ we get $c_1^2$, for $m=4$ $c_1^2(c_1^2 +c_2^2)$, and so on. We remark that onsite terms of the form $b_n |n\rangle \langle n|$ can also be estimated \cite{couplingestimation}. A more thorough analysis of such polynomials in terms of their Gr{\"o}bner basis was recently provided by \cite{paola}.

\begin{figure}
\includegraphics[width=0.45\textwidth]{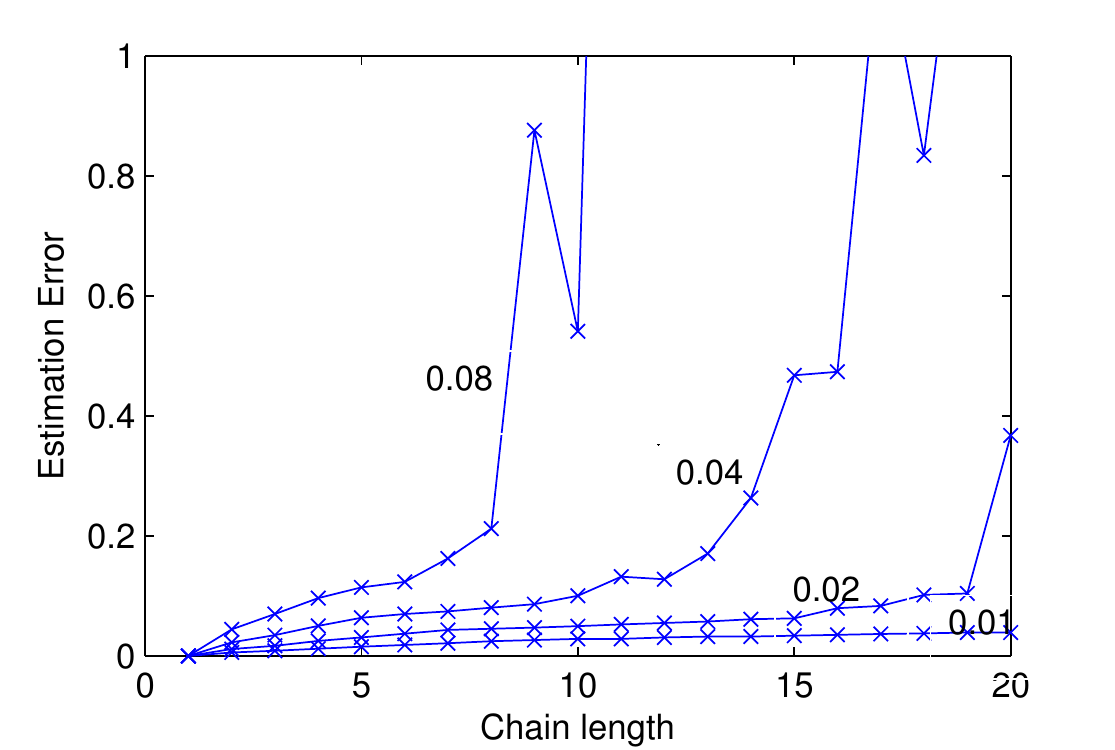}
\caption{\label{numerics}Average error in the estimation of the $c_n$ in Eq.~(1). Shown is $\sum_{n=1}^{N-1}(c_n-\hat{c}_n)^2/(N-1)$ averaged over $1000$ samples for chains of length $N=2,\ldots,20$. The true couplings where taken to be $c_n=1$ and the estimated couplings $\hat{c}_n$ provided by the tomography schemed described in the text. The field parameter was chosen to be large ($f=10$). The four lines correspond to Gaussian noise in the spectral measurements of standard deviations $0.01-0.08$.}	
\end{figure}

\textit{Stability.---}
We numerically analyze the stability of the algorithm with respect to errors in the spectroscopy. Modelling the error as independent Gaussian noise, we find that the error scales weakly with the chain length up to a critical value, after which the estimation scheme fails (see Fig.~\ref{numerics}). For the chains considered, the spectrum is bounded by the interval $[-2,2]$ and the maximum eigenvalue is close to $2$. Thus a standard deviation in the Gaussian noise of $0.04$ corresponds to a percentage of the range of eigenvalues measured. As we can see, the errors must be below that order of magnitude to give useful estimation results for chains of length range up to $N=20$. As a rough argument, the error must be smaller than the smallest difference in eigenenergies, which scales as $~2/N^2$ for isotropic chains. The corresponding critical values $N_c=5,7,10,14$ roughly match the numerical observations. Although in the general scheme the value of the parameter $f$ is not important, for stability it is clear that we want $f$ as large as possible to get a big spectral difference between $H$ and $H'$. This is confirmed by numerics, which scales best when $f>>1$ (we found the error saturates after approximately $a=10$).

\textit{Generalizations.---}
The simple model discussed above can easily be extended to other interesting cases. Firstly, consider Heisenberg spin Hamiltonians of the form
\begin{equation}
	H=\sum_{n=1}^N c_n (XX+YY+\Delta ZZ)_{n,n+1}+\sum_{n=1}^N b_n Z_n.
\end{equation} 
These models conserve the total number of excitations. Considering the sector with one excitation, they become equivalent to the tight-binding models discussed above. If one can either initialize such a system in the first excitation sector, or select the spectral lines corresponding to the first excitation sector using the usual Zeeman effect, then the above protocol becomes applicable.

Likewise, for quadratic Hamiltonians of the form

\begin{equation}
	H= \sum_{n=1}^{N-1} A_{n,n+1} a_n^\dagger a_{n+1} + \frac{1}{2} \sum_{n=1}^{N-1} (B_{n,n+1}a^\dagger_n a^\dagger_{m+1} + h.c.)
\end{equation}
with $A$ Hermitian and $B^T=-\epsilon B$ for fermions ($\epsilon=1$) and bosons ($\epsilon=-1$) we can follow a similar protocol to \cite{quadratic} by applying rank one perturbations $f a_1^\dagger a_1$ and $g a_1^\dagger a_1^\dagger  +h.c.$ followed by spectroscopy.

The finaly generalization is to arbitrary networks.  Because the derivation of Eq.~(\ref{mainequation}) is not specific to the choice of $|1\rangle\langle 1|$ as a field we can consider using using different perturbations. In particular, applying $|\psi\rangle\langle\psi|$ followed by spectroscopy provides us with $|\langle e_k|\psi\rangle|^2$. We choose four different operations $|\psi\rangle=|n\rangle,|m\rangle,(|n\rangle+|m\rangle)/\sqrt{2},(|n\rangle+i|m\rangle)/\sqrt{2}$ followed by spectroscopy. This provides us with $|\langle n |e_k\rangle|^2$, $|\langle m |e_k\rangle|^2$, $|\langle n |e_k\rangle|^2+ |\langle m |e_k\rangle|^2+\langle n|e_k\rangle \langle e_k |m\rangle+\langle m|e_k\rangle \langle e_k |n\rangle$ and $|\langle n |e_k\rangle|^2+ |\langle m |e_k\rangle|^2-i \langle n|e_k\rangle \langle e_k |m\rangle+i\langle m|e_k\rangle \langle e_k |n\rangle$ 
from which $\langle e_k|n\rangle $ and $\langle e_j |m\rangle $ can be estimated. Applying these fields on all pairs $(n,m)$ of the graph would therefore provide all eigenstates and thus the full Hamiltonian. 

But we can do much better: we only need to apply the fields on a subset of the full system, provided that they form an `infecting set' \cite{infecting} (also known as `zero-forcing set'  in graph theory \cite{leslie}) and provided that the couplings are positive. Roughly speaking, it suffices to apply fields to the `surface' of the graph. Under those conditions, the protocol provided in \cite{infecting} combined with the above can be applied to estimate all parameters. We can think of such a set as `marker' particles which respond to external perturbations with a Zeeman shift to provide the parameters of the Hamiltonian.

\textit{Experimental implementations.---} We now consider experimental systems where our Hamiltonian estimation protocol with Zeeman markers can be realized, along with potential applications.

Chains of ions trapped in rf-Paul traps~\cite{Monroe02} are a versatile platform for quantum simulation~\cite{Kim10}. Indeed one can engineer M\o lmer-S\o rensen Hamiltonians~\cite{Sorensen99} of the type $H = \sum_{ij} J_{ij}(S_i^+S_j^- + h.c.)$, and retain only nearest-neighbor interactions via Floquet Hamiltonian engineering~\cite{Ajoy13b} techniques, for instance using laser Stark-shift gradients. Building a versatile quantum simulator relies on a precise characterization of the Hamiltonian between the ions of the chain, which to a good approximation is, $J_{ij} \propto \sum_k \frac{b_{i}^kb_{j}^k}{\mu^2 - \omega_k^2}$ -- a function of coupling  $b^k$ to the collective phonon eigenmodes $\omega_k$ that mediate the spin interactions, and where $\mu$ is the laser detuning~\cite{Porras04}. Measuring the couplings accurately will allow one to experimentally verify and benchmark this calculated expression and identify correction terms given by residual phonon contributions~\cite{Wang12}. Overall such Hamiltonian identification will enable the directed engineering of exotic Hamiltonians and ground states in an ion trap system~\cite{Islam13}. The protocol with Zeeman markers is especially suited since one could prepare and manipulate end states of the chain optically and perform spectroscopy~\cite{Senko14}.

The protocol might also find important application in characterizing spin-based quantum chains, consisting either of nuclear or electron spins. Such chains have been proposed as test-beds for quantum simulation and quantum transport, and as ``wires'' to link distributed quantum registers~\cite{Bose03}. Nuclear chains occur in a variety of natural systems, for example ${}^{19}$F chains in solid crystals of Flouroapatite~\cite{Cappellaro07l}, and ${}^{13}$C spins in certain alkane backbones~\cite{Peng05}. In the latter, the spins are coupled by electron-mediated J couplings, which at zero-field intrinsically has the Heisenberg form, $H = \sum_j J_j\vec{I}_{j}\cdot\vec{I}_{j+1} + BI_{zj}$, where one assumes a weak field $B<J_j$ is applied~\cite{Ledbetter11}. Zeeman markers can be readily applied if the ends of the chain are a different nuclear species.  Spectroscopy of the chain eigenmodes can be achieved by preparing initial states that have support on all the eigenstates -- a simple example being a polarized spin at particular location in an otherwise mixed spin chain -- and subsequent readout. Selective preparation of such states can for instance be done via algorithmic cooling~\cite{Boykin02} or targeted hyerpolarization techniques~\cite{Adams09,Theis16}. Spectroscopic readout can be performed directly at zero-field~\cite{Blanchard13} or by field cycling to higher fields~\cite{Weitekamp83b, Ajoy17b}, each revealing spin transitions from which the Hamiltonian eigenvalues can be reconstructed.

In addition, we envision complementary applications for electron spin chains constructed out of Nitrogen-Vacancy (NV) centers and P1 centers in diamond.  Controlled nitrogen-ion implantation allows the deterministic creation of such spin chains, with spacings under 40nm~\cite{Scarabelli16}, and finite conversion efficiencies determine the ratio of NV and P1 sites. The NV centers can be optically polarized and readout, while P1 centers are not directly addressable at the single spin level. Such chains have found wide interest in quantum information~\cite{Cappellaro11,Yao12} and in environment assisted quantum sensing~\cite{Goldstein11}, where the P1 centers can act to \emph{amplify} the magnetic field sensitivity of the NV centers~\cite{Schaffry11}. The sensitivity gains could be significant -- approaching close to the Heisenberg limit in some protocols~\cite{Goldstein11} -- and allowing a plethora of applications in nanoscale magnetometry~\cite{Lovchinsky16,Suskov13l}. However, in practice the poor characterization of the couplings between ``dark'' P1 spins has been the major obstacle -- a problem that would be exactly addressed by our method. Moreover, since the protocol reveals both the couplings as well as on-site fields, it could enable arrayed quantum sensing using an electron spin chain.
 
\textit{Conclusions.---} The Hamiltonian parameter estimation demonstrated above relies on the ability to actively modify the system through the application of local fields. These do not need to be time-dependent, nor do the measurements need to resolve the dynamics. This paves the way to a stable and general parameter estimation. In some systems one might even get away without having to apply local fields,  by instead engineering two different Hamiltonians $H$ and $H'$ which differ only locally, e.g. by attaching a chemical ligand to a certain atom or by adding/removing particles.

An important result from quantum computing \cite{kempe} shows that in principle, very hard (`QMA-complete') problems can be encoded in spectral properties of simple Hamiltonians. From a broader perspective, it is fascinating to speculate which other dynamical properties can be mapped into spectral ones?

\begin{acknowledgements}
We acknowledge support from the EPSRC Grant No. EP/M01634X/1 and R. Islam for discussions.	
\end{acknowledgements}

\section*{Appendix}
The essential equation of this article is Eq.~(2) which we derive here. Its derivation is elementary but cumbersome and follows standard arguments for Green's functions of rank one perturbations \cite[Chapter~6]{economou} up to Eq.~(\ref{greens}). The remainder of the derivation is analogous to one of Gladwell's inverse problems in vibration \cite[Section~4.5]{gladwell}. For completeness we provide a full derivation.

Consider an eigenvalue $e'$ and eigenvector of $|e'\rangle$ of $H'$. We have the eigenequation
\begin{equation}
H|e'\rangle+f|1\rangle\langle1|e'\rangle=e' |e'\rangle.	
\end{equation}
Express $|e'\rangle$ in the eigenbasis $\{|e_n\rangle\}$ of $H$ with eigenvalues $e_n$ as 
\begin{equation}
|e'\rangle=\sum_{n=1}^N \alpha_n|e_n\rangle,
\label{oldinnew}
\end{equation}
 which yields
\begin{equation}
f|1\rangle\langle1|e'\rangle=\sum_{n=1}^N(e'-e_n)\alpha_n|e_n\rangle.
\end{equation}
Upon multiplication with $\langle e_m|$ we obtain
\begin{equation}
\alpha_m=\frac{f\langle e_m|1\rangle\langle1 |e'\rangle}{(e'-e_m)}	
\end{equation}
where we assumed that the spectrum of $H$ and $H'$  have no overlap (this is true for almost all values of $f$). From Eq.~(\ref{oldinnew}) upon multiplication with $\langle1|$ we arrive at
\begin{equation}
	\langle1|e'\rangle=\sum_{n=1}^N \frac{f\langle e_n|1\rangle\langle 1|e'\rangle}{(e'-e_n)}\langle1|e_n\rangle.
\end{equation}
Since $\langle 1|e'\rangle\neq 0$ \cite{vittorio}, this is equivalent to
\begin{equation}
0=1-\sum_{n=1}^N \frac{f |\langle e_n |1\rangle|^2}{(e'-e_n)}	
\end{equation}
or through expansion with $\prod_{m=1}^N(e'-e_m)$
\begin{equation}
0=\frac{\prod_m (e'-e_m)-\sum_n f |\langle e_n |1\rangle|^2 \prod_{m\neq n}(e'-e_m)}{\prod_m(e'-e_m)}.
\label{greens}	
\end{equation}
This holds for any eigenvalue $e'$ of $H'$. The polynomial in $x$ given by the numerator
\begin{equation}
	P(x)=\prod_{m=1}^N (x-e_m)-\sum_{n=1}^N f |\langle e_n |1\rangle|^2 \prod_{m \neq n} (x-e_m)
\end{equation}
and leading term $x^N$ must therefore factorise as 
\begin{equation}
	P(x)=\prod_n (x-e'_n),
\end{equation}
where $e'_n$ are the eigenvalues of $H'$. We can thus write
\begin{eqnarray}
\lefteqn{1-\sum_{n=1}^N \frac{f |\langle e_n |1\rangle|^2}{(x-e_n)}}\nonumber \\&=&\frac{\prod_m (x-e_m)-\sum_n f |\langle e_n |1\rangle|^2 \prod_{m\neq n}(x-e_m)}{\prod_m(x-e_m)}\nonumber\\
&= &\frac{\prod_n(x-e'_n)}{\prod_m (x-e_m)}.
\end{eqnarray}
Multiply with $(x-e_k)$ to obtain
\begin{equation}
	(x-e_k)-\sum_{n=1}^N \frac{f |\langle e_n |1\rangle|^2(x-e_k)}{(x-e_n)}=\frac{(x-e_k)\prod_n(x-e'_n)}{\prod_m (x-e_m)}
\end{equation}
and perform the limit $x\rightarrow e_k$ such that
\begin{equation}
	-f |\langle e_k |1\rangle|^2=\frac{\prod_n(e_k-e'_n)}{\prod_{m\neq k} (e_k-e_m)}=(e_k-e'_k)\prod_{m\neq k}\frac{(e_k-e'_m)}{(e_k-e_m)}.
\end{equation}
Finally we arrive at 
\begin{equation}
	|\langle e_k|1\rangle |^2=(e_k'-e_k)/f\prod_{m\neq k}\frac{(e_k-e'_m)}{(e_k-e_m)}
\end{equation}

\end{document}